\documentstyle[aps,prd,tighten,eqsecnum]{revtex}
\begin{document}
\thispagestyle{empty}
\preprint{\baselineskip 18pt\vbox{\hbox{SU-4240-564} \hbox{UCI-93-47}
\hbox{\hep-ph/9401209}}}
\vspace{20pt}
\title{ Heavy Quark Solitons : Strangeness and Symmetry Breaking}
\vspace{25pt}
\author{ Arshad Momen and Joseph Schechter}
\address{ Physics Department, Syracuse University, Syracuse,
New York 13244-1130}
\author{ Anand  Subbaraman}
\address{ Physics Department, University of California, Irvine,
California 92717 }
\date{December,1993}
\maketitle
\vspace{35pt}
\begin{abstract}
We discuss the generalization of the Callan-Klebanov model to the case of heavy
quark baryons. The light flavor group is considered to be $SU(3)$ and the limit
of heavy spin symmetry is taken. The presence of the Wess-Zumino-Witten term
permits the neat development of a picture , at the collective level, of a light
diquark bound to a ``heavy" quark with decoupled spin degree of freedom. The
consequences of $SU(3)$ symmetry breaking are discussed in detail. We point
out that the $SU(3)$ mass splittings of the heavy baryons essentially measure
the ``low energy" physics once more and that the comparison with experiment is
satisfactory.
\end{abstract}
\pagebreak
\vfill
\section{Introduction}

A natural method for describing a baryon containing  a single heavy quark in
the ``soliton picture" is to consider the heavy baryon to be a bound state of a
heavy-meson and a light ``baryon as soliton". This was extensively applied by
Callan and Klebanov \cite{CK} and others \cite{others} to the case where the
$K$-meson is considered heavy. A fairly literal transcription of this approach
was given for the charm and bottom baryons too \cite{rho1}. More recently , it
has been recognized that it is necessary to take into account the Isgur-Wise
heavy spin symmetry \cite{spinsym} when dealing with the chiral interactions of
the heavy mesons. This feature was then incorporated , with somewhat different
results, by two groups \cite{manohar,rho2}. In the present paper , we will
study further a possibly simpler method \cite{us} ,
based on an explicit presentation of
the ans\"atz for the ``classical" bound state. It was also noticed in
\cite{us} that the effect of including light vector mesons in the underlying
chiral Lagrangian was important for estimating the semi-classical binding
energy.

The new points here are mainly concerned with generalizing the treatment of
ref.\cite{us} to light $SU(3)$ so as as to be able to treat heavy baryons with
strangeness. A surprising feature is that this generalization actually
simplifies the procedure. The reason is the existence of the
Wess-Zumino-Witten (WZW) term \cite{witten} in the light $SU(3)$ case but not
in the light $SU(2)$ case. We will see that the interplay of the WZW term and
the heavy meson kinetic term gives an important constraint on the allowed
states of the collective Hamiltonian. What emerges is that the collective
Hamiltonian describes a bosonic ( light diquark) rotator in addition to a
decoupled ( in the heavy spin symmetry limit ) spinor representing the heavy
quark. The heavy quark symmetry is then essentially manifest and the entire
treatment is rather simple. It is amusing that, although the underlying
Lagrangian is a theory of mesons, the collective picture looks
quark-like.
Of course this is implicit in \cite{CK}, but here it will be
seen to follow in a
particularly neat way.

We will also use this formalism to discuss the $SU(3)$ mass splittings of the
heavy baryons. It seems natural to do so in the limit of heavy spin symmetry.
This is because the heavy spin splittings vanish as the heavy quark mass,
$M \rightarrow \infty$, in contrast to the $SU(3)$ splittings which remain
finite in
this limit. The physics which is being probed by these mass splittings is very
similar to that determining the mass splittings of the light baryons. In the
latter
case, a treatment \cite{all} based on lowest order perturbation theory is
inadequate, as may be seen by a comparison with the exact diagonalization of
the collective Hamiltonian \cite{Yabu}. It was pointed out \cite{park1} that
second order perturbation theory does provide an adequate approximation to the
exact solution and that is what will be used here. While it would be most
desirable to compare our predictions with data on the bottom baryons,
there is, at present,
sufficient reliable information only for the charmed baryons.
Comparing our results with the likely experimental $J^P  = \frac{1}{2}^+$
states $\Lambda_c,\; \Sigma_c,\;\Xi_c$ and $\Omega_c$, gives values for the
basic coefficients of the collective Hamiltonian which are reasonably close to
those obtained from studies of the light baryon spectrum. We also predict the
mass of another expected $\Xi_c$ state and note that, in the limit of heavy
spin symmetry, it should not mix with the already observed one.

For the reader's convenience, the underlying fields and chiral Lagrangians are
briefly reviewed in section II. Both the Lagrangians with and without vector
fields are given since the form of the $SU(3)$ invariant collective Lagrangian
is the same in each case. In section III, the classical soliton solution
for the light meson fields as well as the heavy meson bound state ans\"atz are
listed.  It is noted that a putative $J^P = \frac{1}{2}^- $ heavy quark baryon
analogous to the $\Lambda(1405)$ is unbound ( in the heavy spin symmetry limit)
in the model with pseudoscalars only. If vectors are included,
such a state might
be weakly bound, but this depends crucially on a poorly known heavy meson-light
vector coupling constant. The discussion of the collective mode quantization
and its interpretation is given in section IV. Furthermore, the low lying
$SU(3)$ multiplets of heavy baryons are identified and discussed. In section V
we introduce the $SU(3)$ symmetry breaking and find its effects on the heavy
baryon mass splittings. We note the fact that $SU(3)$ symmetry breaking among
the heavy mesons has a rather small effect at the collective Hamiltonian
level; it is the light pseudoscalar meson breaking which actually dominates the
heavy baryon splittings.

\section{The Meson Fields and the Chiral Lagrangians}

In this section we briefly summarize the $SU(3)$ chiral Lagrangians under
consideration.The total action consists of  a ``light" part  describing the
first three flavors (namely $u,d,s$) and a ``heavy" part which describes the
``heavy" multiplet $H$ and its interaction with the light sector:

\begin{equation}
\Gamma_{eff} = \Gamma_{light} + \int \; d^4x\; {\cal L}_{heavy} \qquad.
\label{action}
\end{equation}
The relevant light fields belong to the $3\times3$ matrix of pseudoscalars,
$\phi$,
and to the $3\times3$ matrix of vectors, $\rho_\mu$. It is convenient to define
objects which transform simply  under the action of the
chiral group,

\begin{eqnarray}
\xi=\exp{(\frac{i\phi}{F_\pi})},  \qquad U= \xi^2, \nonumber  \\
A_\mu^L = \xi\rho_\mu \xi^\dagger + {i \over {\tilde{g}}}\xi
\partial_\mu\xi^\dagger,\nonumber \\
A_\mu^R = \xi^\dagger\rho_\mu \xi + {i \over {\tilde{g}}}\xi^\dagger
\partial_\mu\xi,
\nonumber \\
F_{\mu\nu} = \partial_\mu\rho_\nu - \partial_\nu\rho_\mu - i{\tilde{g}}
[\rho_\mu,\rho_\nu],
\end{eqnarray}
where $F_\pi \approx$.132 $GeV$ and ${\tilde{g}} \approx 3.93$ for a typical
fit.

   The interactions of the heavy meson fields can be encoded in a compact way
by using the so-called ``Heavy Superfield" \cite{super} which combines  the
heavy pseudoscalar $P'$ and the heavy vector $Q_\mu'$, both moving with a fixed
4-velocity $V_\mu$ :
\begin{equation}
H = {{1-i\gamma_\mu V_\mu}\over 2}(i\gamma_5 P'+ i\gamma_\nu Q'_\nu),
\;\; {\bar H} \equiv \gamma_4 H^\dagger \gamma_4.
\label{1}
\end{equation}
In our conventions the superfield $H$ has the canonical dimension one. It is a
$4\times4$ matrix in the Dirac space and it also carries an unwritten
flavor index for the light quark bound to the heavy quark. The chiral
interactions of $H$ with the light pseudoscalars were discussed in
\cite{Hchi}. The inclusion of the light vector mesons were
given in \cite{Hvec,them}. Here we follow the notations of
ref. \cite{Hvec}.

Using (\ref{1}), ${\cal L}_{heavy}$ can be simply written as,

\begin{equation}
{{{\cal L}_{heavy}} \over {M}} = iV_\mu Tr \left[ H\left( \partial_\mu -
i\alpha{\tilde{g}}\rho_\mu - i(1-\alpha)v_\mu\right){\bar
H}\right]+id\;Tr\left[
H\gamma_\mu\gamma_5 p_\mu{\bar H}\right] +{ic \over{m_v}}\;Tr \left[ H
\gamma_\mu
\gamma_\nu F_{\mu\nu}(\rho){\bar H}\right],
\label{2}
\end{equation}
where $m_v \approx 0.77\; GeV$ is the light vector mass and

\begin{equation}
v_\mu, p_\mu \equiv {i \over 2} (\xi \partial_\mu \xi^\dagger \pm \xi^\dagger
\partial_\mu \xi).
\end{equation}
Furthermore $M$ is the heavy meson mass and $\alpha,c,d$ are dimensionless
coupling constants for the heavy-light interactions. It seems appropriate
not to include terms in (\ref{2}) which are higher order in $\frac{1}{M}$ or
contain more derivatives of the light meson fields.

The action involving the
light pseudoscalar and vector mesons, $\Gamma_{light}$, can be written as the
sum of a  usual piece,

\begin{equation}
\int d^4x \left[ -\frac{F_\pi^2}{8} Tr (\partial_\mu U \partial_\mu U^\dagger)
-
\frac{1}{4} Tr \, \left( F_{\mu \nu}(\rho)F_{\mu \nu}(\rho)\right) -
\frac{m_v^2}{2{\tilde{g}}^2}
Tr \left[ ({\tilde{g}} \rho_\mu - v_\mu )^2 \right]
\right],
\end{equation}
and a piece proportional to the Levi-Civita symbol. The latter is most
conveniently written,using the differential form notation, in terms of the
one-forms $\alpha_\mu \equiv \partial_{\mu}U U^\dagger \rightarrow \alpha $ and
$A_\mu^L \rightarrow A^L$:

\begin{equation}
\Gamma_{WZW} + \int Tr \left[ ic_1(A^L\alpha^3) + c_2(dA^L\alpha A^L-A^L \alpha
dA^L +A^L \alpha A^L\alpha) +c_3\{-2i(A^L)^3\alpha+{1\over{\tilde{g}}}A^L
\alpha
A^L \alpha\}\right],
\label{3}
\end{equation}
where  the Wess-Zumino-Witten term \cite{witten} is  given by,
\begin{equation}
\Gamma_{WZW} = -{i \over{80\pi^2}}\int_{{\cal M}^5} Tr\; (\alpha^5).
\end{equation}
Note that the $c_1,c_2,c_3$ terms in (\ref{3}) perform the function of
stabilizing the Skyrme soliton in this model. More details are given in
\cite{Jain} (wherein ${\tilde{g}}$ is denoted by $g$).

We shall also consider here a simpler light Lagrangian in which the vectors are
absent. ( In this case the constants $\alpha$ and $c$ in (\ref{2}) should also
be set
to zero). We then have the standard $SU(3)$ Skyrme model,

\begin{equation}
\Gamma_{light} = \int d^4x Tr \left( {F_\pi^2 \over 8} \alpha_\mu \alpha_\mu +
{1 \over {32 e^2}} \left[ \alpha_\mu, \alpha_\nu \right]^2 \right) +
\Gamma_{WZW},
\label{skyrme}
\end{equation}
wherein $e$ is the Skyrme constant.

The discussion of $SU(3)$ symmetry breaking terms is deferred to section V.

\section{Baryon States at the Classical Level}

Following the Callan-Klebanov strategy \cite{CK}, we first find the classical
solution
of $\Gamma_{light}$ and then obtain the classical approximation to the
wavefunction in which this ``baryon as soliton" is bound to a heavy meson
( yielding a {\it heavy} hyperon ). To avoid confusion we
remark that whereas the original approach \cite{CK} dealt with a two-flavor
light action
and considered the strange quark as ``heavy", in the present work we are
dealing with a three-flavor light action, considering the strange quark as
light.When one improves the treatment of flavor symmetry breaking for the
$SU(3)$ soliton one obtains results for the strange hyperons of comparable
accuracy to those of the Callan-Klebanov approach.

The ``hedge-hog" ans\"atz for the classical light baryon in the $SU(3)$ case
simply corresponds to embedding the two-flavor ans\"atz as follows :

\begin{equation}
\xi_c =\left( \begin{array}{cc}
\exp{\frac{i}{2}[{\bbox{\hat{x}\cdot\tau}} F(r)}] & 0\cr
0 & 1 \end{array}\right).
\label{skan}
\end{equation}
In a model with vectors present, we have similarly the classical solutions,

\begin{equation}
\rho_{c\,\mu} = \left(\begin{array}{cc}
{1\over\sqrt2}(\omega_{c\,\mu} + \tau^a \rho^a_{c\,\mu})&0\cr
0&1\end{array} \right),
\label{rhoan}
\end{equation}
with ( see \cite{Jain} for example ),
\begin{eqnarray}
\rho^a_{i\,c} = {1 \over {\sqrt 2 {\tilde{g}}r}} \epsilon_{ika} \hat{x}_k G(r),
\qquad \rho^a_{0\,c}=0,\nonumber \\
\omega_{i\,c}=0, \;\;\;\; \omega_{0\,c}= \omega(r).
\end{eqnarray}
The boundary conditions for a finite energy light baryon  are

\begin{eqnarray}
F(0)=-\pi,\;\;\; G(0)=2, \;\;\;\; \omega'(0)= 0,\nonumber \\
F(\infty) = G(\infty) = \omega(\infty) = 0.
\end{eqnarray}

For describing the soliton-heavy meson bound state it is convenient to define
the grand spin ${\bf G}$ as the sum of isospin and the angular momentum :

\begin{equation}
{\bf G}={\bf I}+{\bf J}
\end{equation}
In \cite{CK} it was found that the attractive channel for the ordinary
hyperon, treated as a bound state of the nucleon-as-soliton and the kaon, was
the one with orbital angular momentum $l=1$ and this was combined with the kaon
isospin to give $G=\frac{1}{2}$. In our previous work
\cite{us}, it was shown that the same situation persisted in the heavy
meson-nucleon bound state. In this case it is more intuitive to think of the
heavy meson as being at rest and the soliton bound to it.
When the heavy meson is
at rest $(V_i=0)$ the $4\times4$ matrix heavy superfield ${\bar H}$ given in
(\ref{1}) has non-vanishing elements only in the lower-left $2\times2$
sub-block:
\begin{equation}
{\bar H}=\left(\begin{array}{cc} 0&0\cr
{{\bar H}}^b_{l\,h}&0\end{array}\right).
\label{block}
\end{equation}
The first lower index $l$ of the submatrix ${{\bar H}}^b_{l\,h}$ represents the
spin of the light degrees of
freedom within the heavy meson, while the second lower index $h$ represents
the spin of the heavy quark. The wavefunction for the heavy field is then
written as

\begin{equation}
{{\bar H}}^b_{l,h} =
     \left\{ \begin{array}{cll}
\frac{1}{\sqrt{8\pi M}}
({\bbox{\hat{x}\cdot \tau}})_{bm}\epsilon_{lm}u(r)\chi_h & \mbox{if }& b=1,2\cr
      0 & \mbox{if }& b=3
             \end{array}
     \right..
\label{an}
\end{equation}
This represents an embedding of (3.1) of \cite{us} into the three dimensional
representation of SU(3). The radial wavefunction $u(r)$ is taken in the
classical approximation to be localized at the origin, $ r^2|u(r)|^2 \approx
\delta(r)$. Clearly, this is reasonable at the limit $ M \rightarrow \infty$.
Note that the quantity $\hat{\bf x}$ represents the angular part of the spatial
wave function and the first factor couples it to the isospin index $b$ to give
$ G= \frac{1}{2}$. In turn, this is coupled to the light spin index $l$ with
the Clebsch-Gordan coefficient $\frac{1}{\sqrt 2} \epsilon_{lm}$ to give $G=0$.
Finally , $h$ is left uncoupled (as appropriate to the heavy spin symmetry )to
give  the desired net result $G= \frac{1}{2}$.The two-component heavy quark
spinor $\chi_h$ (which was implicit in (3.1) of \cite{us})
basically carries the heavy spin.

Substituting the ans\"atz (\ref{an}) into ${\cal L}_{heavy}$ in (\ref{2})
yields a classical binding potential $(V_0 = -\int d^3x {\cal L}_{heavy})$
given in (4.3) of \cite{us},

\begin{equation}
V_0 = -{3 \over 2} d F'(0) + \frac{3c}{m_v \tilde{g}} G''(0) - \frac{\alpha
\tilde{g}}{\sqrt
2} \omega(0).
\label{fit1}
\end{equation}
Information about the light-heavy coupling constants $d$ and $c$ can be
obtained
from the semileptonic $D \rightarrow K$ and $ D \rightarrow K^*$ transitions
respectively.
This implies that the first two terms of (\ref{fit1}) are negative,
suggesting that
$V_0$ is also negative. Indeed, taking into account the effect of the quantum
fluctuations on (\ref{fit1}) (see section V of \cite{us} ) as well as estimates
based on semi-leptonic data gives an approximate fit ( see eq.(2.15) of
\cite{sumati}) :

\begin{equation}
d = 0.53,  \qquad c= 1.6, \qquad \alpha =-2.
\label{num}
\end{equation}
This value of $\alpha$ makes the third term of (\ref{fit1}) negative also.
Hence,
the ans\"atz (\ref{an}) leads to attraction. This evidently also holds in the
model
with pseudoscalars only, in which just the first term of (\ref{fit1}) is kept.
We note, however, that the numbers in (\ref{num}) should be regarded as
preliminary in nature.

 Other ans\"atzae for ${\bar H}$ do not necessarily lead to attraction.
Consider, for
example, the possibility,

\begin{eqnarray}
{{\bar H}}^a_{lh} =\left\{ \begin{array}{cll}

\frac{1}{\sqrt{8\pi M}}\epsilon_{la}w(r)\chi_h
       &\mbox{if }&  a=1,2\cr
      0&\mbox{if }& a=3
             \end{array}
     \right..
\label{otan}
\end{eqnarray}
which corresponds to a zero orbital angular momentum state with $ G=
\frac{1}{2}$. This would be a heavy analog of the  $\Lambda(1405) \;J^P =
\frac{1}{2}^-$ hyperon. Substituting (\ref{otan}) into (\ref{2})
gives the binding
potential :
\begin{eqnarray}
V_0(J^P=\frac{1}{2}^-)=
\frac{d}{2} F'(0) - \frac{c}{\tilde{g} m_v} G''(0) - \frac{\alpha \tilde{g}}
{\sqrt 2}
\omega(0),
\label{bad}
\end{eqnarray}
which should be contrasted with (\ref{fit1}). We note that the first two terms,
corresponding to $\pi$ and $\rho$ exchange interactions, are both positive
(repulsive) in this channel while the third term, corresponding to the $\omega$
exchange interaction, is attractive. Evidently, there is no binding in the
model where only pseudoscalars are included. In the full model, using
(\ref{num})
above and (4.5) of \cite {us}, (\ref{bad}) equals -0.10 $GeV$ whereas the bound
state case (\ref{fit1}) equals -1.77 $GeV$ ( to which an important quantum
fluctuation
term should be added ). Hence, the $\frac{1}{2}^-$  channel would seem to be
rather weakly bound as a result of the $\omega$ term just managing to overcome
the $\pi$ and $\rho$ terms. However, the estimate for $\alpha$ is rather
crude and the true situation is not really clear. An experimentally observed
$\frac{1}{2}^-$ heavy baryon near threshold would support the negative sign of
$\alpha$ in (\ref{num}). Note that, as previously mentioned \cite{Hvec}, a
natural notion of light vector meson dominance for the light-heavy interactions
would suggest $\alpha = +1$. Callan and Klebanov \cite{CK} do
find the $\Lambda(1405)$
as a bound state but  their model is different, as they do not assume heavy
quark symmetry and they do not include light vector particles.

\section{ Collective mode Quantization}

In the soliton approach, the particle states with definite rotational and
flavor quantum numbers do not appear until the so-called ``rotational
collective modes" are introduced and the theory is quantized. This is
conveniently done \cite{adkins} by first finding the time independent
parameters which leave the theory invariant. Then those ``collective"
parameters are allowed to depend on time  . Specifically,we set
\begin{eqnarray}
\xi ({\bf x},t) = A(t) \xi_c({\bf x}) A^\dagger (t), \qquad \rho_\mu ({\bf
x},t) = A(t) \rho_{\mu\, c} ( {\bf x}) A^\dagger (t),\nonumber \\
{\bar H} ({\bf x},t) = A(t){\bar H}_c({\bf x}),\qquad  \qquad  \qquad
\label{rot}
\end{eqnarray}
where the ``classical" bound state wavefunction ${\bar H}_c$ is to be taken
from
(\ref{an}). Here $A(t)$ is an $SU(3)$ matrix which acts on the isospin index of
${\bar H}$. $A$ is conventionally considered as a matrix of  angle-type
variables;  generalized angular velocities $\Omega_k$ are defined by

\begin{equation}
A^\dagger {\dot{A}} = \frac{i}{2} \sum_{k=1}^8 \lambda_k \Omega_k,
\end{equation}
where the $\lambda_k$ are the usual Gell-Mann matrices. Now substituting
(\ref{rot})
into the total effective Lagrangian (\ref{action}) and performing a
spatial integration
eventually yields the following collective Lagrangian:

\begin{equation}
L_{coll}= -M_c - ( M+V_0 )P +\frac{ \alpha^2}{2} \sum_{i=1}^3 \Omega_i \Omega_i
+ \frac{\beta^2}{2} \sum_{j=4}^7 \Omega_j \Omega_j - \frac{\sqrt 3}{2} \Omega_8
+\frac{\sqrt 3}{6} \Omega_8 \chi^\dagger \chi P  .
\label{tocol}
\end{equation}
Here, $M_c$ is the classical soliton mass, $\alpha^2$ is the ordinary moment of
inertia and  $\beta^2$ is the ``strange" moment of inertia. Expressions for
these quantities are given in \cite{park2} for the full model including the
vectors and in \cite{Her1} for the minimal $SU(3)$ Skyrme model containing the
pseudoscalars only. The factor $P$ is a projection operator onto the the heavy
baryon subspace of the theory and appears in those terms in the collective
Lagrangian which originate from terms involving the heavy fields. $V_0$ is the
classical binding energy given in (\ref{fit1}).
(Actually the quantum corrections to
$V_0$ discussed in section $V$ of \cite{us} are also important. ) The last
term in (\ref{tocol}) comes from the heavy meson kinetic term
$iV_\mu\, Tr (H \partial_\mu {\bar H})$ ; we have included in it a factor
$\chi^\dagger \chi = \delta_{hh'}$ pertaining to the heavy spin
indices which makes manifest the fact that the
heavy spin index in (\ref{block}) has {\em not } been
summed over in arriving at this term. Thus the heavy spin represents
a dynamical degree of freedom of the collective Lagrangian. The fact that the
particular
value of the heavy spin is not communicated to the soliton variables is a
reflection of the underlying heavy quark symmetry. This means that a term of
the form
\begin{equation}
{\bf \Omega} \cdot \chi^\dagger \mbox{\boldmath $\sigma$}\chi ,
\end{equation}
can not appear in the present model, which represents a substantial difference
from the Callan-Klebanov paper \cite{CK}.
Notice that the the form of $L_{coll}$ is the
same regardless of whether or not the light vector mesons are included.
However, the specific values of the numerical parameters then differ.

The next step is to quantize (\ref{tocol}). The canonical momenta
( for an implicit
parametrization of the matrices $A$) may be taken as :

\begin{equation}
-R_k = \frac{\partial L_{coll}}{\partial  \Omega_k}.
\label{moment}
\end{equation}
For $k=1,2,\cdots,7$, (\ref{moment}) yields true dynamical momenta.
However, for $k=8$ one gets
\begin{equation}
R_8 =\left\{ \begin{array}{cl} \frac{\sqrt 3}{2}, & \mbox{ for light baryons}
\cr
\frac{1}{\sqrt 3}, & \mbox{ for heavy baryons}.\end{array}\right.
\end{equation}
The above numerical difference together with the presence of the the heavy
quark spin degree of freedom are the main differences between the the present
heavy baryon case and the usual light baryon case. Hence we can make use of
the usual quantization \cite{quan} of the $SU(3)$ Skyrme model. An
operator $R_8$ which obeys the canonical commutation relations can be
introduced but we must demand {\it a la Dirac}, the following constraint
on the allowed states $|\qquad \rangle$  of the model :
\begin{equation}
\frac{2}{\sqrt 3}R_8 |\qquad\rangle =\left\{ \begin{array}{cl}
|\qquad\rangle, & \mbox{ for light baryons}
\cr
\frac{2}{3}|\qquad\rangle, & \mbox{ for heavy baryons}.\end{array}\right.
\label{states}
\end{equation}
Then the $R$'s can be seen to obey an $SU(3)$ algebra  $[R_i, R_j ] = -i
f_{i\,j\,k} R_k$. The spatial components ${\bf R}$ (i.e. for $i=1,2,3$) can be
identified as rotation generators for the soliton rotator  while the quantity
$\frac{2}{\sqrt 3}R_8$ in (\ref{states}) is a conventionally normalized
hypercharge
generator. We also define ``left" generators $ L_j = D_{jk} (A) R_k$, where
the adjoint representation matrix $D_{jk}(A)$ can be written as

\begin{equation}
D_{jk}(A) = \frac{1}{2} Tr \left( \lambda_j A \lambda_k A^\dagger \right).
\label{gener}
\end{equation}
They obey a separate $SU(3)$ algebra $[L_i, L_j] = if_{i\,j\,k}L_k$ and can be
identified as $SU(3)$ flavor generators. Note that $\sum_{m=1}^8 L_m L_m =
\sum_{n=1}^8 R_n R_n $. The collective Hamiltonian is

\begin{equation}
H_{coll} = M_{c}+ (M+V_0)P +\frac{1}{2}(\frac{1}{\alpha^2} -
\frac{1}{\beta^2}){\bf R}^2+ \frac{1}{2\beta^2} \sum_{m=1}^{8} R_m^2 -
\frac{1}{2\beta^2} R_8^2,   \nonumber
\end{equation}
which can be written as

\begin{equation}
H_{coll} = M_c + (M+V_0)P + \frac{1}{2} (\frac{1}{\alpha^2} -\frac{1}{\beta^2})
J_s(J_s+1) +\frac{1}{2\beta^2} C_2(SU(3)_L) - \frac{1}{2\beta^2}R_8^2,
\label{massf}
\end{equation}
where $J_s$ is the soliton angular momentum and $C_2(SU(3)_L)$ is the $SU(3)$
quadratic Casimir operator. Irreducible representations of $SU(3)$ may be
specified by a traceless tensor with $p$ symmetric quark type
indices (say,upper) and $q$ symmetric anti-quark type (say, lower) indices.
Then,

\begin{equation}
C_2 (SU(3)) = \frac{1}{3} (p^2+pq+q^2)+(p+q).
\label{casimir}
\end{equation}

It is evident that $H_{coll}$ is $SU(3)$ flavor invariant. For the
{\em light} baryon subspace, the space of the
``angular" wavefunctions is spanned by $SU(3)$ representation matrices
$D^{(\mu)}(A)$, where $\mu$ denotes the irreducible representation under
consideration. With conventional \cite{what} normalization the light baryon
wavefunctions are :
\begin{equation}
\Psi_{light}(\mu, YII_3,JJ_3;A) = (-1)^{J-J_3} \sqrt{dim \;\mu}\;\; D^{
(\mu)*}_{Y,I,I_3;Y_R \, J,\,\,-J_3}(A).
\label{wave1}
\end{equation}
The composite indices were labelled in accordance with the fact that the flavor
generators $L_k$ act on the left composite index while the generators $R_k$,
which include the space rotation ones, act on the right composite index.
It is crucial to note that the constraint (\ref{states}) implies that the
index $Y_R$ in
(\ref{wave1}) must be set equal to $1$. In turn, this implies that only those
irreducible  representations $\{\mu\}$ are allowed which contain a state with
$Y=+1$. Such states, as we will see below, have half odd-integral spin and so
must represent fermions. The rotational spin of the wavefunction (\ref{wave1})
 equals
the iso-spin of the desired state belonging to $\{\mu \}$ which has $Y=1$.

The {\em heavy} baryon wavefunctions may be
constructed analogously. Due to the constraint (\ref{states}), now the angular
wavefunction must have $Y_R= \frac{2}{3}$. In this case, the allowed
$D^{(\mu)}(A)$'s must have integral spin, since $SU(3)$ states with $Y=
\frac{2}{3}$ necessarily have integral iso-spin. To see this, let us consider
a $(p,q)$ tensor component described above which has $n_i$=the number of
$i$th-type quark indices minus the number of $i$th-type
antiquark indices, where $i= u,d,s$. Such a state is labelled by

\begin{eqnarray}
I_3 = \frac{1}{2}(n_u - n_d),\nonumber \\
Y= \frac{1}{3}(n_u+n_d -2n_s).
\label{hyp}
\end{eqnarray}
Substituting $Y=\frac{2}{3}$ into the last equation above gives $n_u+ n_d =
2(1+n_s)$. Subtracting $2n_d$ from both sides finally yields

\begin{equation}
I_3 = 1+n_s - n_d = \mbox{ integer}.
\end{equation}
(If we had substituted $Y=1$ into (\ref{hyp}) we would have obtained $I_3=
\frac{3}{2} + (n_s-n_d) = \frac{1}{2}$ odd integer). The fact that
$D^{(\mu)}(A)$ has integral spin meshes perfectly with the need to include in
the overall heavy baryon wavefunction the Pauli spinor $\chi_h$, which remains
in the collective Lagrangian (\ref{tocol}) ( even though it decouples in
accordance
with the heavy quark symmetry ). Now, we can write the heavy baryon
wavefunction as

\begin{equation}
\Psi_{heavy} (\{\mu\}, Y\,I\,I_3 ; J\, J_3,J_s; A ) = \sum_{h=1}^2 C^{J_s
\,\,\frac{1}{2} \,\,J}_{M_s\,\,h\,\,J_3} (-1)^{(J_s - M_s)} \sqrt{dim \; \mu}
\chi_h D^{ (\mu)*}_{Y,I,I_3;\frac{2}{3},J_s,-M_s} (A),
\label{waveh}
\end{equation}
wherein $J_s$ and $M_s$ are the soliton spin and its $z$-component while the
first factor on the right-hand side is an ordinary $SU(2)$ Clebsch-Gordan
coefficient. Notice that for a given $J_s$ there are two possible values of
the total spin $J = J_s \pm \frac{1}{2}$ which yield degenerate states of
$H_{coll}$. These two states comprise a heavy quark spin symmetry multiplet.
This symmetry is essentially manifest in the present treatment.

It is interesting to observe that, even though the starting model describes
the interactions of heavy and light mesons, at the collective level a picture
more like the quark model emerges; namely, a heavy quark spinor is compounded
with a light diquark wavefunction $D^{(\mu)}(A)$. This picture is not a matter
of choice  but is  forced upon us by the constraint in (\ref{states}). This
constraint results from the presence of both the three flavor
Wess-Zumino-Witten
term (which gives the next to last term in (\ref{tocol}) ) and the heavy field
kinetic
term with the ``classical" ans\"atz (\ref{an}) ( which gives the last term in
(\ref{tocol})).

Now let us apply this approach to the low lying baryons. The simplest $SU(3)$
multiplet with a $Y=\frac{2}{3}$ member is the $\{{\bar{\bf 3}}\}$ ( with
$(p,q)
= (0,1)$). Its $(Y,I)$ content is $\{(\frac{2}{3},0), (-\frac{1}{3},
\frac{1}{2})\}$. Since the $Y= \frac{2}{3}$ state has $I=0$ we conclude that
the
soliton spin $J_s= 0$. Combining this with the heavy quark spinor in
(\ref{waveh})
yields net spin $\frac{1}{2}$ baryons, which are denoted $\{\Lambda_Q,
\Xi_Q ({\bar
{\bf 3}})\}$. The subscript $Q$ indicates that one $s$-quark in the ordinary
hyperon has been replaced by the heavy quark $Q$.

The next simplest $SU(3)$ multiplet with a $Y=\frac{2}{3}$ member is the
$\{{\bf
6}\}$ ( with $(p,q)= (2,0)$). Its $(Y,I)$ content is
$\{(\frac{2}{3},1),(-\frac{1}{3},\frac{1}{2}),(-\frac{4}{3},0)\}$. Since the
$Y=\frac{2}{3}$ state has $I=1$, we conclude that the soliton spin $J_s=1$.
Combining this with the heavy quark spinor in (\ref{waveh}) yields degenerate
multiplets with net spins $J=\frac{1}{2}$ and $\frac{3}{2}$. These are denoted
as $\{\Sigma_Q, \Xi_Q({\bf 6}), \Omega_Q\}$ and $\{\Sigma^*_Q, \Xi_Q^*({\bf 6})
, \Omega_Q^*\}$ respectively.

The fifteen states mentioned are all the low-lying ($s$-wave ) baryon states in
the quark model containing a single heavy quark. In the limit of exact flavor
$SU(3)$ symmetry it is easy to evaluate the mass splittings by acting with
$H_{coll}$ given in (\ref{massf}) on the wavefunction (\ref{waveh}).
 For example, one can readily see that

\begin{equation}
m({\bf 6}, \frac{1}{2}) = m({\bf 6 }, \frac{3}{2}),
\label{equal}
\end{equation}
since \begin{eqnarray} \left( C^{ 1\,\,\frac{1}{2}
\,\,J}_{M-\frac{1}{2}\,\,\frac{1}{2}\,\,M}\right)^2 +
\left( C^{ 1\,\,\frac{1}{2}
\,\,J}_{M+\frac{1}{2}\,\,-\frac{1}{2}\,\,M}\right)^2 =1. \nonumber
\end{eqnarray}
Eq.(\ref{equal}) is just the expression of heavy quark symmetry. In addition,
using
$C_2({\bf {\bar 3}}) = \frac{4}{3}$ and $C_2({\bf6}) = \frac{10}{3}$ from
(\ref{casimir}), we find

\begin{equation}
m({\bf 6}) - m({\bf {\bar3}}) = \frac{1}{\alpha^2} .
\label{boom}
\end{equation}
Similarly treating the light baryons using (\ref{massf}) we find

\begin{equation}
m(\Delta) - m(N) = \frac{3}{2\alpha^2}
\end{equation}
where $\Delta$ and $N$ stand for the usual decuplet and octet baryons. From the
last two formulas we get the structural relation,

\begin{equation}
m({\bf 6 }) - m({\bar {\bf 3}}) = \frac{2}{3} [ m(\Delta) - m(N)].
\label{stru}
\end{equation}

Not surprisingly, (\ref{stru}) agrees with the $SU(2)$ relation given,
e.g. in \cite
{manohar,us}, where it was noted to be reasonably well satisfied experimentally
for $m(\Sigma_c) - m(\Lambda_c)$.

\section{ $SU(3)$ Symmetry Breaking }

In this section we will use perturbation theory to discuss the mass splittings
within the heavy baryon $SU(3)$ multiplets mentioned in the last section. We
restrict ourselves to the limit of degenerate heavy quark spin multiplets. Of
course, the spin splittings (between states of the same flavor) should vanish
as $ M \rightarrow \infty$. In contrast, the flavor splittings within a
multiplet of
given spin should not vanish as $M \rightarrow \infty$ and thus are a
characteristic
feature of the present model.

At the level of the fundamental QCD Lagrangian the flavor splittings are
induced by the light quark mass terms,

\begin{equation}
{\cal L}_{mass} = -\hat{m} {\bar q} {\cal M}q ,
\end{equation}
where $q$ is the column vector of $u,d,s$ quark fields, $\hat{m}=\frac{m_u
+m_d}{2} $ and ${\cal M}$ is a dimensionless, diagonal matrix which can be
expanded
as follows,

\begin{equation}
{\cal M} = y \lambda_3 + T +xS ,
\end{equation}
with  $\lambda_3 = $ diag $(1,-1,0)$, $T=$ diag $(1,1,0)$, and $S=$ diag
$(0,0,1)$. $x$ and $y$ are the quark mass ratios,

\begin{equation}
x=\frac{m_s}{\hat{m}}, \qquad \qquad y=\frac{m_u - m_d}{2\hat{m}}.
\end{equation}
It will be assumed as usual that all the {\em effective} symmetry breaking
 terms are proportional to ${\cal M}$. In ref.
\cite{Big} it was shown that a rather detailed fit to both the light
pseudoscalar and light vector systems required six different terms
- three quark-line rule conserving terms for the pseudoscalars and three
analogous terms
for the vectors. For simplicity, we shall keep here just the dominant term
involving only the pseudoscalars :

\begin{equation}
{\cal L}_{SB}= \delta ' \;\;Tr [ {\cal M} (  U+ U^\dagger - 2 )] + \cdots,
\label{snap}
\end{equation}
where   $\delta ' \approx 4.04 \times 10^{-5} \;\; GeV^4$, $x \approx 31.5$ and
$y\approx \,-0.42$ \cite{Big}. We shall also neglect the small isospin
violation
by assuming $y=0$. Eq. (\ref{snap}) may alternatively be regarded as an
appropriate
symmetry breaking term for the minimal $SU(3)$ Skyrme model with pseudoscalars
only, as given in eq.(\ref{skyrme}). The contribution of (\ref{snap}) to the
collective Lagrangian is obtained by substituting
eq. (\ref{rot}) and the result is,
\begin{equation}
L^{SB}_{coll} = \frac{16 \pi \delta '}{3} \left[ 3+ (x-1)(1-
D_{88}(A))\right] \int r^2 dr ( cos F(r) -1),
\label{lsb}
\end{equation}
where $D_{88}(A)$ is defined in (\ref{gener}). Conventionally \cite{Yabu},
the term
with the overall factor of $3$ is included in $M_c$.

It is reasonable to expect that the effective symmetry breaking terms for the
heavy meson multiplet $H$ (in (\ref{1})) should also influence the heavy baryon
mass splittings. The leading term of this type has the form \cite{SB}:

\begin{equation}
{\cal L}_{\ SB,h} = \epsilon M \left[ tr \left( H \xi {\cal M} \xi {\bar H}
\right) + h.c. \right].
\label{heavysnap}
\end{equation}
The  parameter $\epsilon$ may be obtained in terms of the mass difference
between the strange heavy meson and the non-strange heavy meson,

\begin{equation}
\epsilon = \frac{M_s - M}{2(x-1)}.
\end{equation}
The charmed meson case yields \cite{PDG}

\begin{equation}
m(D_s^{+*}) - m(D^{+*})= m(D_s^+) - m(D^+) = 100 {\rm MeV}
\end{equation}
which is accurate up to 2\%. Evidently, the heavy quark spin prediction,
implicit in (\ref{heavysnap}), works quite well. The contribution of
(\ref{heavysnap}) to the
collective Lagrangian is obtained with the substitutions (\ref{skan}) and
(\ref{an})
followed by a spatial integration :

\begin{equation}
L_{coll}^{hSB} = \frac{2\epsilon}{3} \left[ (2+x) + (1-x)D_{88}(A)\right]
P+\cdots
\label{hsb}
\end{equation}
where the projector $P$ maps into the heavy baryon subspace as usual.

We shall take the sum of terms proportional to $D_{88}(A)$, from (\ref{lsb})
 and
(\ref{hsb}), as our perturbation operator, $H'$ ( which transforms
like $\lambda_8$ in the flavor space ):

\begin{eqnarray}
L' = -H' = -\tau D_{88}(A), \nonumber \\
\tau = \tau_{light} + \tau_{heavy},
\label{break}
\end{eqnarray}
where $\tau_{heavy} = \frac{1}{3} (M_s - M) \approx 0.034 \;\; GeV$.
$\tau_{light}$
depends on the soliton profile $F(r)$, as can be seen from (\ref{lsb}), and is
sensitive in its details to the parameters of the light effective Lagrangian
(e.g. the value of the Skyrme parameter $e$ in (\ref{skyrme}) ).
Typical values for
$\tau_{light}$ are in the $-0.6 \pm 0.2$ $GeV$ range \cite{Yabu,park2,Her1},
 which shows
that the effect of (\ref{heavysnap}) involving the heavy fields
on the heavy baryon mass splittings is in fact rather small.

Before going further, we remark that the earliest treatments \cite{all}
 of the
$SU(3)$ Skyrme model for the light baryons gave predictions which did not
compare well with experiment. There were several reasons for this. The first
is that perturbation thery was carried out only to the first order. Later,
Yabu and Ando \cite{Yabu} showed that the collective Hamiltonian with the
symmetry breaker (\ref{break}) can be diagonalized exactly, by numerical means,
and the results were considerably improved. It was then noted \cite{park1}
that an adequate approximation to the exact solution can be obtained by using
perturbation theory to the second order. The results could be further improved
by taking  a kind of  ``strangeness cranking" \cite{Her1,park2} into account
which had the
effect of increasing the strange moment of inertia $\beta^2$. With these
improvements, quite reasonable predictions for the many baryon octet and
decuplet
mass splittings were obtained. Still, it was necessary to accept a rather
large overall baryon mass (assuming that the well known quantities
like $F_\pi$
take on their experimental values ).More recently, it has been noted
\cite{Holz} that $O(N_c^0)$ corrections  are likely to lower the
overall baryon mass drastically, without modifying the mass splittings.
Keeping these lessons in mind, we will carry out perturbation theory to second
order and focus our attention on mass splittings rather than overall masses.In
fact we will see that the structural relations for the mass splittings require
going beyond first order perturbation theory.

Denoting the matrix elements of (\ref{break}) between the heavy baryon states
in
(\ref{waveh}) by $H_{ab}'$, we compute the mass corrections as :

\begin{equation}
\Delta\,m_a = H_{aa}' - \sum_n \frac{|H_{na}'|^2}{m_n-m_a} + \cdots
\label{pert}
\end{equation}

The possible intermediate states $n$ which can contribute to the sum in (5.11)
are determined from the $SU(3)$ decompositions,
\begin{eqnarray}
{\bar{\bf 3}} \otimes {\bf 8} = {\bar {\bf 3}} \oplus {\bf 6} \oplus {\bar {\bf
15}},\nonumber \\
{\bf 6} \otimes {\bf 8}= {\bar {\bf 3}} \oplus {\bf 6} \oplus {\bar {\bf
15}} \oplus {\bf 24}.
\end{eqnarray}
Now, as we see from (\ref{waveh}), due to the conservation of heavy spin in the
effective theory,
the heavy baryon states are also
labelled by the soliton spin, $J_s$ (which can be called the spin of the light
degree of freedom ). The $\{{\bar {\bf 15}}\}$ has $Y= \frac{2}{3}$ states
with both $I=0$ and $I=1$, so there are two different eigenstates of
(\ref{massf}),
namely - $({\bar {\bf 15}},0)$ and  $({\bar {\bf 15}},1)$. Because $H'$
does not alter the soliton spin, $H_{{\bar{\bf 3}},{\bf6}}' = 0$ and also
$H'$ will have a vanishing matrix element between ${\bar {\bf 3}}$ and
$({\bar {\bf 15}},1)$. However, $H'$ has a non-vanishing matrix element between
${\bar{\bf 3}}$ and  $({\bar {\bf 15}},0)$. The required matrix elements
can be expressed in terms of the $SU(3)$ isoscalar factors by the formula
\cite{SU3} :

\begin{equation}
\langle \{\mu'\},YII_3;JMJ_s|
D_{88}(A)|\{\mu\}, YII_3;JMJ_s\rangle =
\sqrt{\frac{dim \;\mu}{dim \; \mu'}}
\left(\begin{tabular}{cc|c}{\bf8}&$\mu$&$\mu'$\\
$00$&$YI$&$YI$\end{tabular}\right)
\left(\begin{tabular}{cc|c}{\bf8}&$\mu$&
$\mu'$\\
$00$&$\frac{2}{3}J_s$&$\frac{2}{3}J_s$\end{tabular}\right).
\label{matrix}
\end{equation}
The states here correspond to the heavy baryons in (\ref{waveh}).
Note that, due to
the heavy spin conservation, the $SU(2)$ Clebsch-Gordan coefficients in
(\ref{waveh})
do not show up in the final result. The needed $SU(3)$ iso-scalar factors are
given in ref.\cite{CG}. We then find the matrix elements in Table 1.
Lastly, the ``energy denominators" in (\ref{pert}) can be read off from
(\ref{massf})
 to be:

\begin{eqnarray}
m({\bar{\bf 15}},0) - m({\bar{\bf 3}}) = \frac{2}{\beta^2},\nonumber \\
m({\bar{\bf 15}},1) - m({\bf 6}) = \frac{1}{\beta^2}, \nonumber \\
m({\bf 24},1) - m({\bf 6}) = \frac{15}{6\beta^2}.
\end{eqnarray}
Putting things together gives the mass corrections for the low lying baryons
containing a single heavy quark :
\begin{eqnarray}
\Delta \, m(\Lambda_Q ) &=& \frac{\tau}{4} - \frac{9}{160}\tau^2 \beta^2,
\nonumber\\
\Delta \, m(\Xi_Q({\bar{\bf 3}}) ) &=& -\frac{\tau}{8} - \frac{27}{640}
\tau^2 \beta^2,\nonumber \\
\Delta \, m(\Sigma_Q ) &=& \frac{\tau}{10} - \frac{29}{250}\tau^2 \beta^2
+\frac{1}{\alpha^2},\nonumber \\
\Delta \, m(\Xi_Q({\bf 6})) &=& -\frac{\tau}{20} - \frac{123}{2000}\tau^2
\beta^2
+\frac{1}{\alpha^2},\nonumber\\
\Delta \, m(\Omega_Q ) &=& -\frac{\tau}{5} - \frac{3}{125}\tau^2 \beta^2
+\frac{1}{\alpha^2},\nonumber\\
\label{masspred}
\end{eqnarray}
where the $\{{\bf 6}\}$-$\{{\bar{\bf 3}}\}$ splitting of $1/{\alpha^2}$
discussed in sect.$IV$ is also included. There is one additional prediction of
the model. Generally, one would expect the states $\Xi_Q({\bar {\bf
3}})$ and $\Xi_Q({\bf 6})$ to mix under a $\lambda_8$ type perturbation.
However
, because our states also conserve the soliton spin $J_s$, this mixing cannot
occur to any order in $H'$. To obtain the $\Xi({\bf 6})$-$\Xi({\bar{\bf 3}})$
mixing one must include additional terms which take account of the ``hyperfine
interactions".

How well do the  predictions (\ref{masspred}) agree with experiments ?
Considering that
we have worked throughout to the leading order in $M$, it would be best to
test them for the $b$-baryons. However, at present, sufficient data exists only
for the
$c$-baryons. The $J^P = \frac{1}{2}^+$ states will be taken to
have the masses ({\it all in} $GeV$) :

\begin{eqnarray}
m(\Lambda_c) = 2.285 , \qquad m( \Xi_c({\bar{\bf 3}}))=2.470, \nonumber\\
m(\Sigma_c) = 2.453 , \qquad m(\Omega_c) = 2.706 .
\end{eqnarray}
The first three masses (averaging over members of the iso-multiplets where
necessary ) were taken from the particle data Tables \cite{PDG}, while the
$\Omega_c$ mass was taken from \cite{Omega}.Only one $\Xi_c$ state has
apparently been confirmed ; we have assigned it to the  $\{{\bar{\bf 3}}\}$
rather than $\{{\bf 6}\}$. The reason is that the observed state lies very
far from the average of the  $\Sigma_c$ and the
$\Omega_c$ masses. Since the$\{{\bf 6}\}$ is a ``triangular" representation of
$SU(3)$, the Gell-Mann Okubo mass formula  ( which approximately holds in the
light $SU(3)$ Skyrme model even though the second order terms are important
\cite{park1}) does predict equal spacing of the levels, which would be
badly contradicted by the $\{ {\bf 6}\}$ assignment of $\Xi_c(2470)$.

For orientation, let us first examine the predictions of (\ref{masspred})
when second
order  $\tau^2 \beta^2$  terms are excluded. We would then have the relation
between the $\{{\bf 6}\}$ and $\{{\bar {\bf 3}}\}$  splittings :
\begin{eqnarray}
m(\Omega_c) - m(\Sigma_c) = \frac{4}{5} \left[ m(\Xi_c({\bar{\bf 3}})) -
m(\Lambda_c)\right], \nonumber
\end{eqnarray}
which reads numerically as $0.253 = \frac{4}{5}(0.185)$. Clearly, first order
perturbation theory is inadequate. At the second order, we have  three known
mass differences given in terms of three parameters whose range of values are
known from the study of the light $SU(3)$ Skyrme model :
\begin{eqnarray}
m(\Xi_c({\bar{\bf 3}})) - m(\Lambda_c) &=& -\frac{3}{8}\tau +
\frac{9}{640}\tau^2\beta^2, \nonumber\\
m(\Omega_c) - m(\Sigma_c) &=& -\frac{3}{10}\tau +
\frac{23}{250}\tau^2\beta^2, \nonumber\\
m(\Sigma_c) - m(\Lambda_c) &=& -\frac{3}{20}\tau -
\frac{239}{4000}\tau^2\beta^2 +\frac{1}{\alpha^2} .
\end{eqnarray}
{}From these three equations we extract the parameters of the
collective Hamiltonian  :
\begin{equation}
\tau = -0.542\;\;GeV ,\qquad \alpha^2= 6.08 \;\;GeV^{-1} \qquad
\beta^2 = 4.43 \;\;GeV^{-1}.
\label{roughfit}
\end{equation}
We also predict the mass of the other $\Xi_c$ state, belonging to $\{{\bf
6}\}$:
\begin{equation}
m(\Xi_c({\bf 6})) = m(\Sigma_c) - \frac{3}{20}\tau +
\frac{109}{2000}\tau^2\beta^2 = 2.603 \,\,GeV,
\end{equation}
which is not too far from the equal spacing prediction of $2.580\;\; GeV$.
It is in
fact of great interest to compare the parameters given in (\ref{roughfit})
 to those
obtained (see table I of \cite{Her1}) by using a nearly minimal $SU(3)$ Skyrme
model for the light baryons including both Yabu-Ando \cite{Yabu} and
``$K$-cranking" improvements :

\begin{eqnarray}
\tau = -\frac{\gamma}{2} + 0.034 = -0.635\; GeV,\nonumber\\
\alpha^2 = 6.74 \; GeV^{-1}, \qquad \beta^2 = 5.23 \; GeV^{-1}.
\label{lightfit}
\end{eqnarray}
(We have indicated the connection between our parameter $\tau$ and the
parameter $\gamma$ given in \cite{Her1}).
Considering the fact that we are working in the leading $M$ limit and the
general accuracy of the Skyrme approach, the agreement between (\ref{roughfit})
 and
(\ref{lightfit}) is quite encouraging.Note especially that the value of
$\beta^2$ in (\ref{roughfit}) is larger than that which can be gotten without
$K$-cranking . We are persuaded to believe that further
improvement of the present model will be able to provide another useful window
on non-perturbative QCD.

The relatively simple form of the present model should facilitate the
investigation of other issues including the relaxation of the point like
treatment
of the heavy mesons, fine tuning of the $SU(3)$ symmetry breaking and the
inclusion of the ``hyperfine " interactions. It would be interesting to study
the electromagnetic and weak properties of the heavy baryons and to examine
the other channels like $J^P = \frac{1}{2}^-$ in more detail.

{\it Note added}: After this paper was submitted we learned of the recent
interesting papers \cite{int} which discuss other approaches to the
quantization rules for heavy quark baryons .

\section*{Acknowledgements}

We would like to thank Kumar Gupta and Herbert Weigel for interesting
discussions. One of us (A.M.) would also like to thank Dave Besson
and Giancarlo Moneti for helpful
conversations. This work was supported in part  by the U.S. DOE contract
no. DE-FG-02-85ER40231 and NSF grant PHY-9208386.

\vspace{30pt}
{\underline{\bf Table Caption:}}
\begin{itemize}
\item{ Table I : Matrix Elements in (\ref{pert})}
\end{itemize}
\begin{tabular}{|l|c|c|}\hline
 &  $\Lambda_Q$  &  $\Xi_Q({\bar{\bf 3}})$\\ \hline
$H'_{{\bar{\bf 3}},{\bar{\bf 3}}}$ & $\frac{\tau}{4}$ & $-\frac{\tau}{8}$
\\ \hline
$H'_{{\bar{\bf 15}},{\bar{\bf 3}}}$ & $\frac{3\tau}{4\sqrt 5}$ &
$\sqrt{\frac{27}{320}}\tau$ \\ \hline
\end{tabular}
\hspace{2cm}
\begin{tabular}{|l|c|c|c|}\hline
& $\Sigma_Q$ & $\Xi_Q({\bf 6})$ & $\Omega_Q$ \\ \hline
$H'_{{\bf6 },{\bf 6}}$ & $\frac{\tau}{10}$ & $-\frac{\tau}{20}$ &
$-\frac{\tau}{5}$ \\ \hline
$H'_{{\bar{\bf15}},{\bf 6}}$ & $\frac{\tau}{\sqrt 10}$ &
${\sqrt{\frac{3}{80}}}\tau$ & 0 \\ \hline
$H'_{{\bf 24 },{\bf 6}}$ &  $\frac{\tau}{5}$ & $\frac{\sqrt{6}\tau}{10}$ &
$\frac{\sqrt{6}\tau}{10}$ \\ \hline
\end {tabular}
\centerline{Table I}
\end{document}